\documentclass[universe,article,accept,pdftex,moreauthors]{Definitions/mdpi} 
\usepackage{graphicx}
\usepackage{dcolumn}
\usepackage{latexsym,amsfonts}
\usepackage{textcomp}
\usepackage{bm}
\firstpage{1} 
\pubvolume{10}
\issuenum{11}
\articlenumber{415}
\pubyear{2024}
\copyrightyear{2024}
\externaleditor{Academic Editor: Máté Csanád}
\datereceived{18 September 2024} 
\daterevised{3 November 2024} 
\dateaccepted{4 November 2024} 
\datepublished{6 November 2024} 
\hreflink{https://doi.org/10.3390/\linebreak universe10110415} 

\Title{Neutron Beta Decay and Exact Conservation of Charged Weak Hadronic Vector Current in the Standard Model}

\TitleCitation{Neutron Beta Decay and Exact Conservation of Charged Weak Hadronic Vector Current in the Standard Model}

\Author{{Derar Altarawneh} 
 $^{1,}$*,  Roman~H\"ollwieser $^{2}$\orcidA{} and  Markus Wellenzohn $^{3,4}$}

\AuthorNames{Derar Altarawneh,  Roman~H\"ollwieser and Markus Wellenzohn}

\AuthorCitation{Altarawneh, D.; H\"ollwieser,~R.; Wellenzohn, M.}

\address{%
$^{1}$ \quad Department of Applied Physics, Tafila Technical University, Tafila 66110, Jordan\\
$^{2}$ \quad Department of Theoretical Physics, Faculty of Mathematics and Natural Sciences, Wuppertal University, Gau{\ss}stra{\ss}e 20, 42119 Wuppertal, Germany 
\\
$^{3}$ \quad Department of Engineering, Applied Electronics and Technical Informatics, {FH} 
 Campus Wien, \linebreak University of Applied Sciences Vienna, 1100 Vienna, Austria \\
$^{4}$ \quad Research Center IT-Security, Department of Engineering, FH~Campus~Wien, University of Applied Sciences Vienna, 1100 Vienna, Austria}

\corres{{Correspondence:} 
 derar@ttu.edu.jo}

\abstract{We investigate the reliability of the hypothesis of exact conservation
of the charged weak hadronic vector current in neutron
$\beta^-$-decay with a polarized neutron and an unpolarized proton and
electron. 
We calculate the contributions of the phenomenological term
responsible for Exact Conservation of the charged weak hadronic Vector
Current (or the ECVC effect) in neutron $\beta^-$-decay, even for
different masses of the neutron and proton, 
to the correlation coefficients,
together with a complete set of contributions of scalar and tensor
interactions beyond the Standard Model (SM). We argue that if the total
contributions of scalar and tensor interactions beyond the SM 
fail to reconcile the experimental data on the correlation
coefficients with the contributions of the ECVC effect, one may
conclude that the charged weak hadronic vector current is not
conserved in the hadronic transitions of weak processes with different
masses of incoming and outgoing hadrons.} 



\begin{document}

\section{Introduction}
\label{sec:introduction}

Nowadays, it is generally accepted that neutron $\beta^-$-decay is
a perfect laboratory for tests of the Standard Model (SM)
\cite{Abele2008, Nico2009, PERC, Konrad2015, Abele2016, Gudkov2006, 
Ivanov2013, Ivanov2017b}. The theoretical analysis of 
neutron $\beta^-$-decay at the level of $10^{-3}$, including the
radiative corrections of order $O(\alpha/\pi)$, where $\alpha$ is the
fine--structure constant~\cite{PDG2016}, and corrections of order
$1/M$, caused by the weak magnetism and proton recoil, where $M$ is an
averaged nucleon mass, provide a theoretical background at the level
of $10^{-3}$~\cite{Gudkov2006,Ivanov2013,Ivanov2017b} for experimental
searches of interactions beyond the SM at the level of $10^{-4}$
\cite{Ivanov2013,Ivanov2017b}. The calculation of the main
contributions to neutron $\beta^-$--deca,y as well as
corrections of order $10^{-3}$, has been carried out in the $V - A$
theory of weak interactions and QED, where $V$ and $A$ are charged
weak hadronic vector $V^{(+)}_{\mu}(x)$ and axial $A^{(+)}_{\mu}(x)$
currents, respectively. The contribution of the charged weak hadronic
vector current to the matrix element of the hadronic $n \to p$
transition has been calculated at the standard assumption that local
conservation of $V^{(+)}_{\mu}(x)$, i.e., a vanishing divergence
$\partial^{\mu}V^{(+)}_{\mu}(x) = 0$, can be violated only by isospin
breaking~\cite{Marshak1969}. Another possibility for the
contribution of the charged weak hadronic vector current to the matrix
element of the hadronic $n \to p$ transition has been proposed by
Leitner {et al.}~\cite{Leitner2006}. According to Leitner {et
  al.}~\cite{Leitner2006}, the matrix element $\langle p(\vec{k}_p,
\sigma_p)|V^{(+)}_{\mu}(0)|n(\vec{k}_n, \sigma_n)\rangle$ of the
hadronic $n \to p$ transition is defined by
\newpage
\vspace{-6pt}

\begin{adjustwidth}{-\extralength}{0cm}
\begin{eqnarray}\label{eq:1}
\hspace{-0.3in}\langle
p(k_p,\sigma_p)|V^{(+)}_{\mu}(0)|n(k_n,\sigma_n)\rangle =
\bar{u}_p(k_p,\sigma_p)\Big(\Big(\gamma_{\mu} - \frac{q_{\mu}
  \hat{q}}{q^2}\Big)\,F_1(q^2) + \frac{i\sigma_{\mu\nu}q^{\nu}}{2
  M}\,F_2(q^2) \Big) u_n(k_n,\sigma_n),
\end{eqnarray}
\end{adjustwidth}
where $F_1(q^2)$ and $F_2(q^2)$ are form factors dependent on the
squared 4--momentum transferred $q^2 = (k_p - k_n)^2$,
$\bar{u}_p(k_p,\sigma_p)$ and $ u_n(k_n,\sigma_n)$ are the Dirac wave
functions of the free proton and neutron, $\gamma_{\mu}$ and
$\sigma_{\mu\nu} = \frac{i}{2}(\gamma_{\mu}\gamma_{\nu} -
\gamma_{\nu}\gamma_{\mu})$ are the Dirac matrices and $\hat{q} =
\gamma_{\nu}q^{\nu}$, $M = (m_n + m_p)/2$, where $m_n$ and $m_p$ are
masses of the neutron and proton, respectively~\cite{PDG2016}.  The
matrix element Equation~(\ref{eq:1}) obeys the condition
\begin{eqnarray}\label{eq:2}
\hspace{-0.3in}q^{\mu} \langle
p(k_p,\sigma_p)|V^{(+)}_{\mu}(0)|n(k_n,\sigma_n)\rangle = 0
\end{eqnarray}
or $\langle
p(k_p,\sigma_p)|\partial^{\mu}V^{(+)}_{\mu}(0)|n(k_n,\sigma_n)\rangle
= 0$ even for different masses of the proton and neutron.  As has been
shown by Ivanov~\cite{Ivanov2017c,Ivanov2017d}, the terms with the
Lorentz structures $\gamma_{\mu}$, $q_{\mu} \hat{q}$ and
$i\,\sigma_{\mu\nu}q^{\nu}$ are induced by the first-class currents
\cite{Weinberg1958}. According to Ivanov~\cite{Ivanov2017d}, the
matrix element of the hadronic $n \to p$ transition, caused by the
charged weak hadronic vector current, has the following general
structure
\vspace{-9pt}
\begin{adjustwidth}{-\extralength}{0cm}
\begin{eqnarray}\label{eq:3}
\hspace{-0.3in}&&\langle
p(k_p,\sigma_p)|V^{(+)}_{\mu}(0)|n(k_n,\sigma_n)\rangle =
\bar{u}_p(k_p,\sigma_p)\Big(\gamma_{\mu} \,F_1(q^2) +
\frac{i\sigma_{\mu\nu}q^{\nu}}{2 M}\,F_2(q^2) +
\frac{q_{\mu}\hat{q}}{M^2}\,F_4(q^2) + \frac{q_{\mu}}{M}\,F_3(q^2)
\Big) u_n(k_n,\sigma_n).
\end{eqnarray}
\end{adjustwidth}
where $F_j(q^2)$ with $j = 1,2,3,4$ are form factors. The first three
terms with the Lorentz structures $\gamma_{\mu}$,
$i\,\sigma_{\mu\nu}q^{\nu}/M$ and $q_{\mu} \hat{q}/M^2$ are induced by
the first-class current~\cite{Ivanov2017c,Ivanov2017d}, whereas the
term with the Lorentz structure $q_{\mu}/M$ is a phenomenological
contribution of the second-class one~\cite{Ivanov2017d}. According to
Weinberg~\cite{Weinberg1958}, the contribution of the second-class
currents should vanish. This gives $F_3(q^2) = 0$.  Thus, the general
structure of the matrix element of the hadronic $n \to p$ transition,
caused by the contributions of the first-class current only, is
\vspace{-9pt}
\begin{adjustwidth}{-\extralength}{0cm}
\begin{eqnarray}\label{eq:4}
\hspace{-0.3in}&&\langle
p(k_p,\sigma_p)|V^{(+)}_{\mu}(0)|n(k_n,\sigma_n)\rangle =
\bar{u}_p(k_p,\sigma_p)\Big(\gamma_{\mu} \,F_1(q^2) +
\frac{i\sigma_{\mu\nu}q^{\nu}}{2 M}\,F_2(q^2) +
\frac{q_{\mu}\hat{q}}{M^2}\,F_4(q^2) \Big) u_n(k_n,\sigma_n).
\end{eqnarray}
\end{adjustwidth}

We would like to mention that for the first time, the term with the
Lorentz structure $q_{\mu}\hat{q}/M^2$ was introduced by Berman
and Sirlin~\cite{Berman1962} as a part of the Lorentz-invariant
structure of the nucleon vector form factor, with initial and final
nucleons on-mass and off-mass shell, respectively (see Equation~(8) of
Ref.~\cite{Berman1962}).

Of course, the form factor $F_4(q^2)$ is a new phenomenological
function of $q^2$ with respect to the form factors $F_1(q^2)$ and
$F_2(q^2)$, which are usually used for the description of the hadronic
$n \longleftrightarrow p$ transitions. Moreover, it is not obvious that
$F_4(q^2) = - (M^2/q^2)\,F_1(q^2)$. Such a relation we may call a
realization of the hypothesis of Exact Conservation of the charged
weak hadronic Vector Current (or the ECVC effect~\cite{Ivanov2017a}).
For the first time, the contribution of the term $-
F_1(q^2)\,q_{\mu}\hat{q}/q^2$ or the contribution of the ECVC effect
has been analyzed by Leitner {et al.} ~\cite{Leitner2006} in the
quasi-elastic neutrino--neutron scattering $\nu_{\ell} + n \to p +
\ell^-$, where $(\nu_{\ell},\ell^-)$ are the neutrino and charged lepton
with lepton flavours $\ell = e, \mu$, and so on. Such an analysis of
the quasi-elastic neutrino--neutron scattering with the ECVC effect
has been carried out by Leitner {et al.}~\cite{Leitner2006} in
connection with neutrino production of the resonance $P_{33}(1232)$
or the $\Delta$-resonance, i.e.,  $\nu_{\ell} + N \to \Delta +
\ell^-$.

The use of the ECVC effect in the quasi-elastic neutrino--neutron
scattering in connection with the neutrino production of the resonance
$P_{33}(1232)$ or the $\Delta$-resonance is not a surprise. Indeed,
it is well known that exact conservation of the charged weak hadronic
vector current or the ECVC hypothesis is usually used for the
analysis of neutrino production of baryon resonances by a nucleon $N$,
i.e.,  in the reactions $\nu_{\ell} + N \to R + \ell^-$,
\cite{Dufner1968,Schreiner1973a,Schreiner1973b,Singh1998,Paschos2005,Paschos2006a,Paschos2006b} (see also
\cite{Leitner2006}). Here, $R$ is a baryon resonance such as
$P_{33}(1232)$, $P_{11}(1440)$, and so on
\cite{Paschos2005,Paschos2006a,Paschos2006b}, where the numbers in
parentheses define resonance masses $m_R$ measured in ${\rm MeV}$.  In
this case, the matrix elements $\langle
R|\partial^{\mu}V^{(\pm)}_{\mu}(0)|N\rangle$ vanish in spite of the
fact that masses $m_N$ and $m_R$ are different, i.e., $m_N \neq m_R$, even in
the limit of isospin symmetry.  Recently, the ECVC effect or the
contribution of the phenomenological term $-
F_1(q^2)\,q_{\mu}\hat{q}/q^2$ to the cross section for the inverse
$\beta$-decay $\bar{\nu}_e + p \to n + e^+$ has been analyzed by
Ankowski~\cite{Ankowski2016}. 

It is very likely that scattering processes
like quasi-elastic neutrino--neutron scattering and inverse
beta decay are not sensitive enough to the contribution of the ECVC
effect. Indeed, the contribution of the term $-
F_1(q^2)\,q_{\mu}\hat{q}/q^2$ to the matrix element of the hadronic $n
\to p$ transition of the quasi-elastic scattering $\nu_e + n \to p +
e^-$ and the inverse $\beta^-$-decay $\bar{\nu}_e + p \to n + e^+$ is
equal to $\mp F_1(q^2)\,m_e \Delta/q^2$, where $m_e = 0.5110\,{\rm
  MeV}$ and $\Delta$ are the electron mass and the mass--difference
$\Delta = m_n - m_p = 1.2934\,{\rm MeV}$, respectively, and $q^2 =
m^2_e - 2E_{\nu}E_e(1 - \beta\,\cos\vartheta_{e\nu})$
\cite{Ivanov2013a}. As we show in Appendix~\ref{appendixa}, the cross sections for
the quasi-elastic electron neutrino--neutron scattering and for the
inverse $\beta$-decay are not sensitive to the contributions of the
ECVC effect (see Figure~\,\ref{fig:fig2}). For the illustration of the
sensitivity of the cross sections for the quasi-elastic electron
neutrino--neutron scattering and for the inverse $\beta$-decay to the
ECVC effect, we calculate in Appendix~\ref{appendixa}, the cross sections for the
neutrino and antineutrino energy regions $2\,{\rm MeV} \le E_{\nu} \le
8\,{\rm MeV}$ and $2\,{\rm MeV} \le E_{\bar{\nu}} \le 8\,{\rm MeV}$
\cite{Ivanov2013a}, where $E_{\nu}$ and $E_{\bar{\nu}}$ are the
neutrino and antineutrino energy, respectively. In these neutrino and
antineutrino energy regions, the form factor $F_1(q^2)$ can be replaced
by unity, i.e., $F_1(q^2) = 1$~\cite{Ivanov2013a}. 

In Figure~\,\ref{fig:fig2} we plot the relative contributions of the ECVC
effect to the cross sections for the reactions under
consideration. One may see that the contributions of the ECVC effect
decrease with the neutrino and antineutrino energies. The contribution
of the ECVC effect to the cross section for the quasi-elastic
electron neutrino--neutron scattering makes up about $0.7\,\%$ at
$E_{\nu} \simeq 2\,{\rm MeV}$ and one order of magnitude smaller at
$E_{\nu} \simeq 8\,{\rm MeV}$. In turn, the contribution of the ECVC
effect to the cross section for the inverse $\beta$-decay is of about
$3\,\%$ at $E_{\bar{\nu}} \simeq 2\,{\rm MeV}$ and two orders of
magnitude smaller at $E_{\bar{\nu}} \simeq 8\,{\rm MeV}$. It is
important to emphasize that the contribution of the ECVC effect to the
cross section for the inverse $\beta$-decay is negative. This implies
that such a contribution may only increase a deficit of positron
\cite{Mention2013}. Since the cross section for the inverse
$\beta$-decay should be averaged over the reactor electron
antineutrino--energy spectrum, which has a maximum at $E_{\bar{\nu}}
\simeq 4\,{\rm MeV}$, one may expect an increase of the deficit of the
reactor electron antineutrinos by about $0.5\,\%$. Although this is
important as a hint for a search of light sterile neutrinos with a
mass $m_{\nu_s} \sim 1\,{\rm eV}$
\cite{Abazajian2012,Giunti2016,Giunti2017}, the contribution of the
ECVC effect to the yield of positrons $Y_{e^+}$ is smaller than the
experimental error bars $Y_{e^+} = 0.943(23)$~\cite{Mention2013}. So
one may argue that the cross section for the inverse $\beta$-decay is
insensitive to the contributions of the ECVC effect.

As has been shown in~\cite{Ivanov2017a}, neutron $\beta^-$-decay
and, namely, the neutron lifetime has turned out to be extremely
sensitive to the ECVC effect. Indeed, the contribution of the ECVC
effect changes the neutron lifetime by $\Delta \tau_n = 76.4\,{\rm
  s}$. Together with the theoretical value $\tau^{(\rm SM)}_n =
879.6(1.1)\,{\rm s}$, calculated by Ivanov { et al.}
\cite{Ivanov2013}, the ECVC effect increases the neutron lifetime equal to
$\tau^{(\rm eff)}_n = 950(1.1)\,{\rm s}$, which disagrees strongly
with the world averaged value $\bar{\tau}_n = 880.2(1.0)\,{\rm s}$
\cite{PDG2016} and recent experimental value $\tau^{(\exp)}_n =
880.2(1.2)\,{\rm s}$~\cite{Arzumanov2015}. As
has been shown in~\cite{Ivanov2017a}, since in the SM there are no
interactions, which can cancel such a huge contribution of the ECVC
effect, only scalar and tensor interactions beyond the SM can be used
to reconcile the ECVC effect with the world averaged value
$\bar{\tau}_n = 880.2(1.0)\,{\rm s}$ of the neutron lifetime
\cite{PDG2016}. In a linear approximation the contributions of
scalar and tensor interactions beyond the SM reduce themselves to the
neutron lifetime in the form of the Fierz interference term
$b_F$. Cancelling the contribution of the ECVC effect one obtains $b_F
= 0.1219(12)$~\cite{Ivanov2017a}. As has been shown in
\cite{Ivanov2017a}, taking into account the contributions of scalar
and tensor interactions beyond the SM without any approximation, one
may cancel the contribution of the ECVC effect without fixing the
value of the Fierz interference term. The latter is because the
equation which reduces $\tau^{(\rm eff)}_n$ to $\bar{\tau}_n$, is a
quadratic algebraical equation depending on four complex parameters,
which are phenomenological coupling constants of scalar and tensor
interactions beyond the SM~\cite{Ivanov2013}. Thus, the analysis of
the contribution of the ECVC effect to the neutron lifetime, carried
out in~\cite{Ivanov2017a}, has shown the form factor $F_4(q^2)$ cannot
be equal to $- (M^2/q^2)\,F_1(q^2)$ in the SM. Moreover, in comparison
with the results, obtained in~\cite{Ivanov2013,Ivanov2017b}, the
contribution of the term $F_4(q^2)\,q_{\mu}\hat{q}/M^2$ to the neutron
lifetime, calculated in the SM without contributions of interactions
beyond the SM, should be of order $O(m_e\Delta/M^2)$ or even smaller.

Nevertheless, as has been pointed out by Ivanov { et al.}
\cite{Ivanov2017a}, in order to understand a validity of the
hypothesis of exact conservation of the charged weak hadronic vector
current or the ECVC effect, corresponding to a vanishing matrix
element $\langle h'|\partial^{\mu}V^{(+)}_{\mu}(0)|h\rangle$, i.e.,
$\langle h'|\partial^{\mu}V^{(+)}_{\mu}(0)|h\rangle = 0$, of a
hadronic $h \to h'$ transition for different masses of incoming $h$
and outgoing $h'$ hadrons, and to estimate the values of the
phenomenological coupling constants of scalar and tensor interactions
beyond the SM one has to investigate experimentally all asymmetries of
neutron $\beta^-$-decay, expressed in terms of the electron and
proton energy and angular distributions, obtained in
\cite{Ivanov2013,Ivanov2017b} and supplemented by the contributions of
the ECVC effect and scalar and tensor interactions beyond the SM,
calculated without any approximation.

This paper is addressed to the analysis of the contributions of the
ECVC effect, i.e., the contributions of the term $-
F_1(q^2)\,q_{\mu}\hat{q}/q^2$, to the matrix element $\langle
p|V^{(+)}_{\mu}(0)|n\rangle$ of the hadronic $n \to p$ transition
providing a vanishing matrix element $\langle
p|\partial^{\mu}V^{(+)}_{\mu}(0)|n\rangle$ even for different masses
of the neutron and proton. We calculate the contributions of the ECVC
effect to the correlation coefficients of the electron energy and
angular distributions of neutron $\beta^-$-decay with a polarized
neutron and unpolarized proton and electron.

The paper is organized as follows. 
In Section~\ref{sec:spectrum} we define the electron energy and its 
angular distribution in neutron $\beta^-$-decay $n \to p + e^- + \bar{\nu}_e$.
In Section~\ref{sec:asymmetry} we
calculate the correlation coefficient ${\cal A}(E_e)$, defining the
electron asymmetry, with the contribution of the ECVC effect and the
Fierz interference term in the linear approximation for scalar and
tensor interactions beyond the SM. In Section~\ref{sec:conclusion} we
discuss the obtained results, compare the theoretical correlation
coefficient ${\cal A}(E_e)$ with the available experimental data. The
correlation coefficient ${\cal A}(E_e)$ contains the contributions of
(i) the correlation coefficient $A^{(\rm SM)}(E_e)$, calculated at the
level of $10^{-3}$ in the SM~\cite{Ivanov2013}, (ii) the ECVC effect
and (iii) the Fierz interference term.

We propose for new experimental investigations of the contributions
of the ECVC effect together with the contributions of scalar and
tensor interactions beyond the SM, to use the effective expressions for the correlation coefficients of the neutron
$\beta^-$-decay. These are given in Equations~(\ref{eq:14}) and (\ref{eq:18}) for a polarized neutron, unpolarized proton and electron and neutron lifetime $\tau^{(\rm eff)}_n$. They take into account the contributions of the SM, calculated at the level of $10^{-3}$, the ECVC effect and scalar and tensor interactions beyond the SM, calculated to linear approximation for vector and axial vector interactions beyond the SM~\cite{Ivanov2013,Ivanov2017a}. In 
Appendix~\ref{appendixa}, we calculate the cross sections for the quasi-elastic
electron neutrino--neutron scattering and for the inverse
$\beta$-decay by taking into account the contributions of the ECVC
effect in the laboratory frame and in the non-relativistic limit for
the outgoing hadron. We show that the cross sections of the reactions
under consideration are not sensitive to the contributions of the
ECVC effect.

\section{Electron Energy and Angular Distribution}
\label{sec:spectrum}

Following~\cite{Ivanov2017a}, the amplitude of the neutron
$\beta^-$-decay $n \to p + e^- + \bar{\nu}_e$ is given in the standard
form by
\vspace{-10pt}
\begin{adjustwidth}{-\extralength}{0cm}
\begin{eqnarray}\label{eq:5}
\hspace{-0.3in} M(n \to p\,e^-\,\bar{\nu}_e) = -
\frac{G_F}{\sqrt{2}}\,V_{ud}\,\langle p(\vec{k}_p,
\sigma_p)|J^{(+)}_{\mu}(0)|n(\vec{k}_n,
\sigma_n)\rangle\,\Big[\bar{u}_e(\vec{k}_e, \sigma_e)\gamma^{\mu}(1 -
  \gamma^5)\, v_{\nu}(\vec{k}, + \frac{1}{2})\Big],
\end{eqnarray}
\end{adjustwidth}
where $G_F$ and $V_{ud}$ are the Fermi weak coupling constant and the
Cabibbo--Kobayashi--Maskawa (CKM) matrix element~\cite{PDG2016},
$\bar{u}_e(\vec{k}_e,\sigma_e)$ and $v_{\nu}(\vec{k}_{\nu}, +
\frac{1}{2})$ are the Dirac wave functions of the free electron and
antineutrino with 3-momenta $\vec{k}_e$ and $\vec{k}_{\nu}$ and
polarizations $\sigma_e = \pm 1$ and $+ \frac{1}{2}$
\cite{Ivanov2013,Ivanov2014}, respectively, and $\gamma^5$ is the Dirac
matrix.  The matrix element of the charged weak hadronic $V - A$
current $\langle p(\vec{k}_p, \sigma_p)|J^{(+)}_{\mu}(0)|n(\vec{k}_n,
\sigma_n)\rangle$ is given by~\cite{Ivanov2017a}
\vspace{-6pt}
\begin{adjustwidth}{-\extralength}{0cm}
\begin{eqnarray}\label{eq:6}
\hspace{-0.3in}\langle p(\vec{k}_p, \sigma_p)|J^{(+)}_{\mu}(0)
|n(\vec{k}_n, \sigma_n)\rangle = \bar{u}_p(\vec{k}_p,
\sigma_p)\Big[\Big(\gamma_{\mu} - \frac{q_{\mu}\hat{q}}{q^2}\Big) +
  \frac{\kappa}{2M}\,i\,\sigma_{\mu\nu}q^{\nu} + \lambda\,\Big( -
  \frac{2 M \,q_{\mu}}{q^2 - m^2_{\pi}} +
  \gamma_{\mu}\Big)\,\gamma^5\Big]\,u_n(\vec{k}_n, \sigma_n).
\end{eqnarray}
\end{adjustwidth}
 where $q = k_p - k_n$ is a 4-momentum transferred and $-
 q_{\mu}\hat{q}/q^2$ is the phenomenological term responsible for the
 ECVC in neutron $\beta^-$-decay.  Then, the term $\kappa/2 M$,
 where $M = (m_n + m_p)/2$ is the averaged nucleon mass, defines the
 contribution of the weak magnetism, where $\kappa = \kappa_p -
 \kappa_n = 3.7058$ is the isovector anomalous magnetic moment of the
 nucleon, measured in nuclear magneton~\cite{PDG2016}. The
 contribution of the axial current is given by the last term in
 Equation~(\ref{eq:6}), where $\lambda = - 1.2750(9)$ is the axial coupling
 constant~\cite{Abele2008} (see also~\cite{Ivanov2013,Ivanov2014}) and
 $m_{\pi}$ is the charged pion mass~\cite{PDG2016}. In the limit
 $m_{\pi} \to 0$ (or in the chiral limit)
~\cite{Adler1968,DeAlfaro1973}, the matrix element Equation~(\ref{eq:6})
 obeys the constraint $q^{\mu} \langle p(\vec{k}_p,
 \sigma_p)|J^{(+)}_{\mu}(0) |n(\vec{k}_n, \sigma_n)\rangle = 0$, even
 for different masses of the neutron and proton.

Using the results obtained in~\cite{Ivanov2013,Ivanov2017a}, and
skipping standard intermediate calculations, we arrive at the
electron energy and angular distribution for the neutron
$\beta^-$-decay with a polarized neutron and unpolarized proton and
electron
\vspace{-6pt}
\begin{adjustwidth}{-\extralength}{0cm}
\begin{eqnarray}\label{eq:7}
\hspace{-0.3in}&&\frac{d^5 \lambda_n(E_e, \vec{k}_e, \vec{k}_{\nu},
  \vec{\xi}_n)}{dE_e d\Omega_e d\Omega_{\nu}} = (1 + 3
\lambda^2)\,\frac{G^2_F|V_{ud}|^2}{32\pi^5}\,(E_0 - E_e)^2
\,\sqrt{E^2_e - m^2_e}\, E_e\,F(E_e, Z = 1)\,\zeta(E_e)\nonumber\\
\hspace{-0.3in}&&\times\,\Big\{1 + \Big[b_F - \frac{1}{1 +
    3\lambda^2}\,\Big(\frac{2 m_e\Delta }{q^2} - \frac{m_e
    E_e\Delta^2}{q^4}\Big)\Big]\,\frac{m_e}{E_e} + \Big[- \frac{1}{1 +
    3\lambda^2}\,\frac{m^2_e \Delta^2}{q^4} +
  a(E_e)\Big]\,\frac{\vec{k}_e\cdot \vec{k}_{\nu}}{E_e
  E_{\nu}}\nonumber\\
\hspace{-0.3in}&& + A(E_e)\,\frac{\vec{\xi}_n\cdot
  \vec{k}_e}{E_e} + \Big[- b_{ST}\,\frac{m_e}{E_e} - \frac{1}{2}\,(A_0
  + B_0)\,\frac{m^2_e \Delta}{q^2 E_e} + B(E_e)\Big]\,
\frac{\vec{\xi}_n\cdot \vec{k}_{\nu}}{E_{\nu}} +
K_n(E_e)\,\frac{(\vec{\xi}_n\cdot \vec{k}_e)(\vec{k}_e\cdot
  \vec{k}_{\nu})}{E^2_e E_{\nu}}\nonumber\\
\hspace{-0.3in}&&+ Q_n(E_e)\,\frac{(\vec{\xi}_n\cdot
  \vec{k}_{\nu})(\vec{k}_e\cdot \vec{k}_{\nu})}{ E_e E^2_{\nu}} +
\Big[- \frac{1}{2}\,(A_0 + B_0)\,\frac{ \alpha\, m^2_e \Delta}{q^2
    k_e} + D(E_e)\Big]\,\frac{\vec{\xi}_n\cdot
  (\vec{k}_e\times \vec{k}_{\nu})}{E_e E_{\nu}} -
3\,a_0\,\frac{E_e}{M}\,\Big(\frac{(\vec{k}_e\cdot
  \vec{k}_{\nu})^2}{E^2_e E^2_{\nu}} -
\frac{1}{3}\,\frac{k^2_e}{E^2_e}\,\Big) \Big\},
\end{eqnarray}
\end{adjustwidth}
where $\Delta = m_n - m_p$ and $E_{\nu} = E_0 - E_e$ with $E_0 =
(m^2_n - m^2_p + m^2_e)/2m_n = 1.2927\,{\rm MeV}$ is the end-point
energy of the electron energy spectrum, and $q^2 = m^2_e +
2\,k_e\cdot k_{\nu} = m^2_e + 2E_e E_{\nu} - 2\,\vec{k}_e\cdot
\vec{k}_{\nu}$. All terms depending on $\Delta$ are caused by the ECVC
effect. The correlation coefficients $a_0$, $A_0$ and $B_0$ are equal
to~\cite{Abele2008}
\begin{eqnarray}\label{eq:8}
\hspace{-0.3in}a_0 = \frac{1 - \lambda^2}{1 + 3 \lambda^2}\;,\; A_0 =
- 2\,\frac{\lambda(1 + \lambda)}{1 + 3 \lambda^2}\;,\;B_0 =
- 2\,\frac{\lambda(1 - \lambda)}{1 + 3 \lambda^2}.
\end{eqnarray}
They are calculated to leading order in the large nucleon mass
expansion.  Then, $\vec{\xi}_n$ is a unit polarization vector of the
neutron. The correlation coefficients $\zeta(E_e)$, $a(E_e)$, $A(E_e)$,
and so on, have been calculated in~\cite{Ivanov2013} (see also
\cite{Gudkov2006}) within the SM at the level $10^{-3}$ by taking into
account the $1/M$ corrections, caused by the weak magnetism and proton
recoil, and radiative corrections to order $O(\alpha/\pi)$. The
contributions of interactions beyond the SM are calculated to linear
approximation (see Appendix G of Ref.~\cite{Ivanov2013}) and denoted by
$b_F$, which is the Fierz interference term, and $b_{ST}$. The
correlation coefficients $b_F$ and $b_{ST}$ are defined by~\cite{Ivanov2013}
\begin{eqnarray}\label{eq:9}
\hspace{-0.3in}b_F &=& \frac{1}{1 + 3\lambda^2}\,\Big({\rm Re}(C_S -
\bar{C}_S) + 3\lambda\,{\rm Re}(C_T - \bar{C}_T)\Big),\nonumber\\
\hspace{-0.3in}b_{\rm ST} &=& \frac{1}{1 + 3\lambda^2}\,\Big(\lambda\,{\rm
  Re}(C_S - \bar{C}_S) +(1 - 2\lambda)\,{\rm Re}(C_T - \bar{C}_T)\Big),
\end{eqnarray}
where $C_S$, $\bar{C}_S$, $C_T$ and $\bar{C}_T$ are phenomenological
coupling constants of scalar and tensor interactions beyond the SM
\cite{Ivanov2013}.  The contribution of the ECVC effect to the
correlation coefficient $D(E_e)$ is caused by the distortion of the
electron wave function in the Coulomb field of the proton
\cite{Jackson1958,Jackson1958b,Konopinski1966} (see also~\cite{Ivanov2014a,Ivanov2017b,Ivanov:2018uuk,Ivanov:2018vit,Ivanov:2018vmz}).

\section{Electron Asymmetry \boldmath{$A_{\exp}(E_e)$}}
\label{sec:asymmetry}

The electron asymmetry, caused by the correlations between the
electron 3-momentum and neutron spin, is defined by the
electron energy spectrum and angular distribution Equation~(\ref{eq:7})
integrated over the antineutrino 3-momentum $\vec{k}_{\nu}$
\cite{Ivanov2013}. Having integrated over directions of
$\vec{k}_{\nu}$ we get
\begin{adjustwidth}{-\extralength}{0cm}
\begin{eqnarray}\label{eq:10}
\hspace{-0.3in}&&\frac{d^3 \lambda_n(E_e, \vec{k}_e,
  \vec{\xi}_n)}{dE_e d\Omega_e} = (1 + 3
\lambda^2)\,\frac{G^2_F|V_{ud}|^2}{8\pi^4}\,(E_0 - E_e)^2
\,\sqrt{E^2_e - m^2_e}\, E_e\,F(E_e, Z = 1)\,\zeta(E_e)\,\Big\{1 +
\Big[b_F - \frac{1}{1 + 3\lambda^2}\nonumber\\
\hspace{-0.3in}&&\times\,\frac{m_e}{\beta
    E_e}\,\frac{\Delta}{2 E_{\nu}}\, \Big(1 -
\frac{\Delta}{4 E_{\nu}}\Big)\,{\ell n}\Big(\frac{m^2_e + 2E_e
  E_{\nu}(1 + \beta)}{m^2_e + 2E_e E_{\nu}(1 - \beta)}\Big) -
\frac{1}{1 + 3\lambda^2}\,\frac{\Delta}{2 E_{\nu}}\,\frac{m^3_e
  \Delta}{(m^2_e + 2E_e E_{\nu})^2 - 4E^2_e E^2_{\nu}
  \beta^2}\Big]\,\frac{m_e }{E_e}\nonumber\\
\hspace{-0.3in}&& + \Big[A^{(\rm SM)}(E_e) - (A_0 + B_0)\,\frac{m^2_e
    \Delta}{8\beta E^3_e}\, \Big(\frac{m^2_e + 2 E_e E_{\nu}}{2\beta^2
    E^2_{\nu}}\,{\ell n}\Big(\frac{m^2_e + 2E_e E_{\nu}(1 +
    \beta)}{m^2_e + 2E_e E_{\nu}(1 - \beta)}\Big) - \frac{2 E_e}{\beta
    E_{\nu}}\Big)\Big]\,\frac{\vec{\xi}_n\cdot \vec{k}_e}{E_e}\Big\}.
\end{eqnarray}
\end{adjustwidth}

The correlation coefficient $A^{(\rm SM)}(E_e)$ is fully defined by
the contributions of the SM interactions, including the $1/M$
corrections, caused by the weak magnetism and proton recoil {\cite{Wilkinson1982}}, 
and radiative corrections of order
$O(\alpha/\pi)$. It is equal to~\cite{Ivanov2013}
\begin{eqnarray}\label{eq:11}
\hspace{-0.3in}A^{(\rm SM)}(E_e) = A^{(\rm W)}(E_e)\,\Big(1 +
\frac{\alpha}{\pi}\,f_n(E_e)\Big),
\end{eqnarray}
where the correlation coefficient $A^{(\rm W)}(E_e)$ has been
calculated by Wilkinson~\cite{Wilkinson1982} (see also Equation~(20) of
Ref.~\cite{Ivanov2013}). The function $(\alpha/\pi)\,f_n(E_e)$
describes the radiative corrections~\cite{Shann1971} (see also
\cite{Gudkov2006,Ivanov2013}). Following~\cite{Ivanov2013},
the expression of the electron asymmetry $A_{\exp}(E_e)$ is given by
\begin{eqnarray}\label{eq:12}
\hspace{-0.3in}A_{\exp}(E_e) = \frac{N^{+}(E_e) -
  N^{-}(E_e)}{N^{+}(E_e) + N^{-}(E_e)} = \frac{1}{2}\,{\cal A}(E_e)\, P \beta\,(\cos\theta_1 + \cos\theta_2),
\end{eqnarray}
where $N^{\pm}(E_e)$ are the numbers of events of the emission of the
electron forward $(+)$ and backward $(-)$ with respect to the neutron
spin into the solid angle $\Delta \Omega_{12} = 2\pi (\cos\theta_1 -
\cos\theta_2)$ with $0 \le \varphi \le 2\pi$ and $\theta_1 \le
\theta_e \le \theta_2$. Then, $P = |\vec{\xi}_n| \le 1$ is the neutron
polarization and ${\cal A}(E_e)$ is the correlation coefficient, taking
into account the contribution of the ECVC effect and of the Fierz
interference term. It is equal to
\begin{adjustwidth}{-\extralength}{0cm}
\begin{eqnarray}\label{eq:13}
\hspace{-0.3in}&&{\cal A}(E_e)=\Big\{ A^{(\rm SM)}(E_e) - (A_0 +
B_0)\,\frac{m^2_e \Delta}{8\beta E^3_e}\Big[\frac{m^2_e + 2 E_e
    E_{\nu}}{2\beta^2 E^2_{\nu}}\,{\ell n}\Big(\frac{m^2_e + 2E_e
    E_{\nu}(1 + \beta)}{m^2_e + 2E_e E_{\nu}(1 - \beta)}\Big) -
  \frac{2 E_e}{\beta E_{\nu}}\Big]\Big\}\Big\{ 1 + \Big[b_F -
  \frac{1}{1 + 3\lambda^2}\nonumber\\
\hspace{-0.3in}&&\times\,\frac{m_e}{\beta E_e}\,\frac{\Delta}{2
  E_{\nu}}\, \Big(1 - \frac{\Delta}{4 E_{\nu}}\Big)\,{\ell
  n}\Big(\frac{m^2_e + 2E_e E_{\nu}(1 + \beta)}{m^2_e + 2E_e E_{\nu}(1
  - \beta)}\Big) - \frac{1}{1 + 3\lambda^2}\,\frac{\Delta}{2
  E_{\nu}}\,\frac{m^3_e \Delta}{(m^2_e + 2E_e E_{\nu})^2 - 4E^2_e
  E^2_{\nu} \beta^2}\Big]\,\frac{m_e }{E_e}\Big\}^{-1}.
\end{eqnarray}
\end{adjustwidth}

\begin{figure}[H]
\includegraphics[width=.49\textwidth]{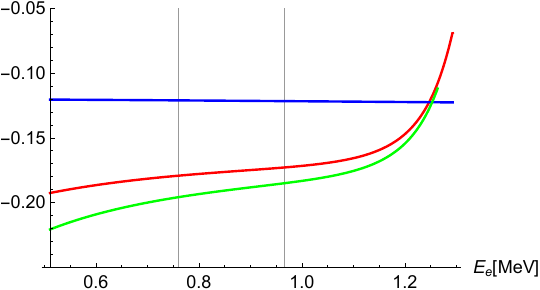}
\includegraphics[width=.49\textwidth]{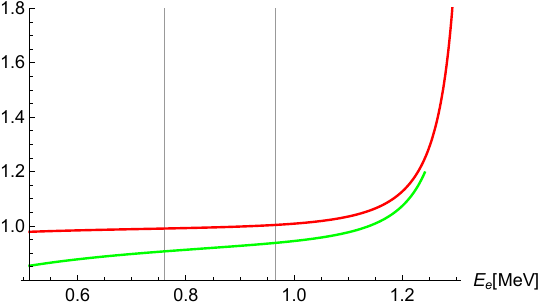}
  \caption{(\textbf{Left}) The correlation coefficient ${\cal A}(E_e)$, given
    by Equation~(\ref{eq:13}) and calculated in the electron energy region
    $m_e \le E_e \le E_0$ for $b_F = 0.1219$ (red) and $b_F = 0$
    (green), and the correlation coefficient $A^{(\rm SM)}(E_e)$
    (blue). (\textbf{Right}) The denominator of the correlation coefficient
    ${\cal A}(E_e)$, plotted in the electron energy region $m_e \le
    E_e \le E_0$ with $b_F = 0.1219$ (red) and $b_F = 0$
    (green).}
\label{fig:fig1}
\end{figure}

\begin{figure}[H]
\includegraphics[width=.49\textwidth]{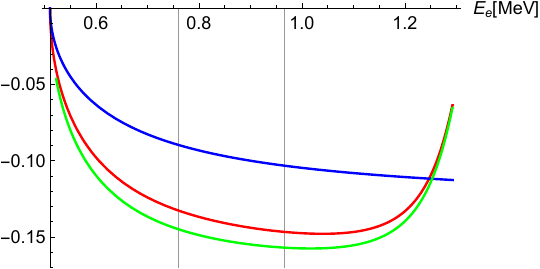}
  \caption{The correlation coefficient $\beta {\cal
      A}(E_e)$, given by Equation~(\ref{eq:13}) and calculated in the
    electron energy region $m_e \le E_e \le E_0$ for $b_F = 0.1219$
    (red) and $b_F = 0$ (green), and the correlation coefficient
    $\beta A^{(\rm SM)}(E_e)$ (blue). The interval between lines $E_e
    = 0.761\,{\rm MeV}$ and $E_e = 0.966\,{\rm MeV}$ defines the
    experimental electron energy region.}
\label{fig:fig1a}
\end{figure}

In Figure~\,\ref{fig:fig1}, we plot the correlation coefficients ${\cal A}(E_e)$,
$A^{(\rm SM)}(E_e)$ in the electron energy region $m_e \le E_e
\le E_0$ (left), and the denominator of the correlation coefficient ${\cal A}(E_e)$ for the Fierz interference term $b_F = 0.1219$ and $b_F = 0$,
respectively (right). In Figure~\,\ref{fig:fig1a}, we show $\beta {\cal A}(E_e)$ for $b_F = 0.1219$ and $b_F = 0$,
respectively, and $\beta A^{(\rm SM)}(E_e)$. The vertical lines constrain the experimental electron energy region $0.761\,{\rm MeV} \le E_e \le 0.966\,{\rm
  MeV}$~\cite{Ivanov2013}.
\vspace{-6pt}
We discuss the results of the theoretical analysis of the electron asymmetry in
comparison with the experimental data in Section \ref{sec:conclusion}.

\section{Discussions and Proposals}
\label{sec:conclusion}

We have continued the analysis of the reliability of the
hypothesis of exact conservation of the charged weak hadronic vector
current in the framework of neutron $\beta^-$-decay, even for different
masses of the neutron and proton, which we have begun in
\cite{Ivanov2017a}.  The contributions of the phenomenological term in
the matrix element of the hadronic $n \to p$ transition, which is responsible
for the Exact Conservation of the charged weak hadronic Vector Current, are referred to as the ECVC effect. We have calculated the contributions of
the ECVC effect together with the contributions of scalar and tensor
interactions beyond the SM, taken in the linear approximation, to the
correlation coefficients of the electron energy and angular
distributions of neutron $\beta^-$-decay with a polarized neutron
and unpolarized proton and electron. We have analyzed the validity of the ECVC hypothesis using example of the electron asymmetry, which is caused by
correlations of the electron 3-momentum and neutron spin.  
In Figures~\,\ref{fig:fig1} and~\ref{fig:fig1a}, we plot the correlation coefficient ${\cal
  A}(E_e)$ for $b_F = 0.1219(12)$ (red line), for $b_F = 0$ (green
line) and also the correlation coefficient $A^{(\rm SM)}(E_e)$ (blue
line), defined by Equation~(\ref{eq:11}). One may see that the correlation
coefficient $A^{(\rm SM)}(E_e)$ agrees well with the experimental
values of the correlation coefficient $A_0$: (i) $A^{(\exp)}_0 = -
0.11933(34)$~\cite{Abele2008}, (ii) $A^{(\exp)}_0 = - 0.11996(58)$
\cite{Abele2013}, (iii) $A^{(\exp)}_0 = - 0.11966 \pm 0.00089^{+
  0.00123}_{-0 .00140}$~\cite{Mendenhall2013}, and (iv) $A^{(\exp)}_0 =
- 0.11832(78)$~\cite{Abele2014}. As has been shown in
\cite{Ivanov2013}, the deviations of $A^{(\rm SM)}(E_e)$ from $A_0$ are
of order $10^{-3}$. This explains a weak energy dependence of $A^{(\rm
  SM)}(E_e)$ on the electron energy $E_e$.  Unfortunately, the energy
dependence of the correlation coefficient ${\cal A}(E_e)$, corrected
by the contribution of the ECVC effect and the Fierz interference
term, differs crucially from that of the correlation coefficient
$A^{(\rm SM)}(E_e)$. This also means that the correlation coefficient
${\cal A}(E_e)$, given by Equation~(\ref{eq:13}), is unable to reproduce the
experimental data of the electron asymmetry, giving the experimental
values of the correlation coefficients $A_0$~\cite{Abele2008,Abele2013,Mendenhall2013,Abele2014}.

At this point, we do not want to argue immediately that the obtained
result means that (i) the ECVC hypothesis is not valid, even in the SM
with scalar and tensor interactions beyond the SM; nor that (ii) the charged
weak hadronic vector current is not conserved in the hadronic $h \to
h'$ transition with different masses of incoming $h$ and outgoing $h'$
hadrons in the sense of a vanishing matrix element $\langle
h'|\partial^{\mu}V^{(+)}_{\mu}(0)|h\rangle = 0$. Hence, we may argue that the
account for the ECVC effect together with the contributions of scalar
and tensor interactions beyond the SM calculated in the linear
approximation does not support the ECVC hypothesis in the neutron
$\beta^-$-decay, at least using example of the electron asymmetry.

As a result, in order to avoid the problem of invalidity of the
ECVC hypothesis with scalar and tensor interactions beyond the SM and
the experimental data, we propose to take into account a complete set of
contributions of scalar and tensor interactions beyond the SM, which
include also the quadratic terms~\cite{Ivanov2013}. Using the results,
obtained in~\cite{Ivanov2013}, we get the following expressions for
the neutron lifetime~\cite{Ivanov2017a} and correlation coefficients
\begin{adjustwidth}{-\extralength}{0cm}
\begin{equation}
\begin{aligned}
\begin{array}{lllll}\label{eq:14}
\hspace{-0.3in}&&\frac{1}{\tau^{(\rm eff)}_n} =\frac{1}{\tau^{(\rm
    SM)}_n}\Big(1 + \frac{1}{2}\,\frac{1}{1 +
  3\lambda^2}\,\Big(|C_S|^2 + |\bar{C}_S|^2 + 3|C_T|^2 +
3|\bar{C}_T|^2\Big) + \frac{\Delta f_n}{f_n} +
b_F\,\Big\langle\frac{m_e}{E_e}\Big\rangle_{\rm SM}\Big),\\
\hspace{-0.3in}&&a_{\rm eff}(E_e, \vec{k}_e\cdot \vec{k}_{\nu}) =
\Big(a(E_e) - \frac{1}{1 + 3\lambda^2}\,\frac{m^2_e \Delta^2}{q^4} +
\frac{1}{1 + 3\lambda^2}\frac{1}{2}\Big(|C_T|^2 + |\bar{C}_T|^2 -
|C_S|^2 - |\bar{C}_S|^2\Big)\Big)/Y(E_e),\\
\hspace{-0.3in}&&A_{\rm eff}(E_e, \vec{k}_e\cdot \vec{k}_{\nu}) = \Big(A(E_e) - \frac{1}{1 +
      3\lambda^2}{\rm Re}\Big(2 C_T \bar{C}^*_T + (C_S
      \bar{C}^*_T + \bar{C}_S C^*_T)\Big)/Y(E_e),\\
\hspace{-0.3in}&&B_{\rm eff}(E_e, \vec{k}_e\cdot \vec{k}_{\nu}) =
\Big(B(E_e) - \frac{1}{2}\,(A_0 + B_0)\,\frac{m^2_e \Delta}{q^2 E_e} -
b_{ST}\,\frac{m_e}{E_e}- \frac{1}{1 + 3\lambda^2}{\rm Re}\Big(2 C_T
\bar{C}^*_T - (C_S \bar{C}^*_T + \bar{C}_S C^*_T)\Big)/Y(E_e),\\
\hspace{-0.3in}&&D_{\rm eff}(E_e, \vec{k}_e\cdot \vec{k}_{\nu})
= \Big(D(E_e) + \frac{2\,{\rm Im}\lambda}{1 + 3\lambda^2} +
\frac{1}{1 + 3\lambda^2}\,\frac{1}{2}\,{\rm Im}\Big(C_S C^*_T +
\bar{C}_S \bar{C}^*_T\Big)\Big)/Y(E_e),
\end{array}
\end{aligned}
\end{equation}
\end{adjustwidth}
with a common denominator
\vspace{-6pt}
\begin{adjustwidth}{-\extralength}{0cm}
\begin{eqnarray}\label{eq:den14}
\hspace{-0.3in}&&Y(E_e)=1 + \frac{1}{2}\,\frac{1}{1 + 3\lambda^2}\,\Big(|C_S|^2 + |\bar{C}_S|^2 + 3|C_T|^2 + 3|\bar{C}_T|^2\Big) + \Big[b_F - \frac{1}{1 +   3\lambda^2}\,\Big(\frac{2 m_e\Delta }{q^2} - \frac{m_e    E_e\Delta^2}{q^4}\Big)\Big]\,\frac{m_e}{E_e},
\end{eqnarray}
\end{adjustwidth}
where $\langle m_e/E_e\rangle_{\rm SM} = 0.6556$
\cite{Ivanov2017a,Ivanov2013}, calculated for the electron energy
density Equation~(D.59) of Ref.~\cite{Ivanov2013}.  The correlation
coefficients $a(E_e)$, $A(E_e)$, $B(E_e)$ and $D(E_e)$ are calculated
in the SM by taking into account the $1/M$ corrections, caused by the
weak magnetism and proton recoil, and radiative corrections of order
$O(\alpha/\pi)$~\cite{Ivanov2013}. Then, $\Delta f_n$ is the
phase-space factor of neutron $\beta^-$-decay, caused by the ECVC
effect. It is given by
\vspace{-6pt}
\begin{adjustwidth}{-\extralength}{0cm}
\begin{eqnarray}\label{eq:15}
\hspace{-0.3in}&&\Delta f_n =  \frac{1}{1 +
  3\,\lambda^2}\int^{E_0}_{m_e}dE_e\,k_e\, (E_0 - E_e)^2\, F(E_e, Z =
1)\int \frac{d\Omega_{e\nu}}{4\pi}\,\Big\{ - \frac{2\,m^2_e\,\Delta
   }{\displaystyle m^2_e + 2 \,E_e (E_0 - E_e) -
    2\,\vec{k}_e\cdot \vec{k}_{\nu}}\nonumber\\
\hspace{-0.3in}&&+ \frac{m^2_e \Delta^2 }{\displaystyle\Big( m^2_e + 2
  \,E_e (E_0 - E_e) - 2\,\vec{k}_e\cdot
  \vec{k}_{\nu}\Big)^2}\,\Big(E_e - \frac{\vec{k}_e\cdot
  \vec{k}_{\nu}}{E_{\nu}}\Big)\Big\} =
\int^{E_0}_{m_e}dE_e\,\sqrt{E^2_e- m^2_e}\,(E_0 - E_e)^2\,F(E_e, Z =
1)\\
\hspace{-0.3in}&&\times\,\Big\{- \frac{1}{1 +
  3\lambda^2}\,\frac{m^2_e}{\beta E_e}\,\frac{\Delta}{2 E_{\nu}}\,
\Big(1 - \frac{\Delta}{4 E_{\nu}}\Big)\,{\ell n}\Big(\frac{m^2_e +
  2E_e E_{\nu}(1 + \beta)}{m^2_e + 2E_e E_{\nu}(1 -
  \beta)}\Big) - \frac{1}{1 + 3\lambda^2}\,\frac{\Delta}{2
  E_{\nu}}\,\frac{m^4_e \Delta}{(m^2_e + 2E_e E_{\nu})^2 - 4E^2_e
  E^2_{\nu} \beta^2}\Big\}\nonumber
\end{eqnarray}
\end{adjustwidth}
and is equal to $\Delta f_n = - 4.887\times 10^{-3}\,{\rm MeV^5}$
\cite{Ivanov2017a}, where $f_n = 6.116\times 10^{-2}\,{\rm MeV^5} $ is
the phase-space factor of neutron $\beta^-$-decay, calculated to
order $O(1/M)$ and $O(\alpha/\pi)$ caused by the contributions of the
weak magnetism and proton recoil and the radiative corrections,
respectively~\cite{Ivanov2013}. As we have shown in
\cite{Ivanov2017a}, the contribution of the term $\Delta f_n = -
4.887\times 10^{-3}\,{\rm MeV^5}$, caused by the ECVC effect, changes
the value of the neutron lifetime by $\Delta \tau_n = (- \Delta f_n/(f_n + \Delta f_n))\,\tau^{(\rm SM)}_n = 76.4\,{\rm
  s}$. Thus, fitting the neutron lifetime $\tau^{(\rm eff)}_n =
\tau^{(\rm SM)}_n$ at the level of current experimental accuracy
$1.2\times 10^{-3}$ or one standard deviation~\cite{Ivanov2017a},
where $\tau^{(\rm SM)}_n = 879.6(1.1)\,{\rm s}$, calculated in
\cite{Ivanov2013}, agrees well with the world averaged value \mbox{$\tau_n =
880.2(1.0)\,{\rm s}$~\cite{PDG2016}} and the experimental value
$\tau^{(\exp)}_n = 880.2(1.2)\,{\rm s}$, we~get
\begin{eqnarray}\label{eq:16}
\frac{1}{2}\,\frac{1}{1 + 3\lambda^2}\,\Big(|C_S|^2 + |\bar{C}_S|^2 +
3|C_T|^2 + 3|\bar{C}_T|^2\Big) = - \frac{\Delta f_n}{f_n} -
b_F\,\Big\langle\frac{m_e}{E_e}\Big\rangle_{\rm SM}.
\end{eqnarray}

Neglecting the contributions of the quadratic terms, we arrive at the
equation~\cite{Ivanov2017a}
\begin{eqnarray}\label{eq:17}
\frac{\Delta f_n}{f_n} +
b_F\,\Big\langle\frac{m_e}{E_e}\Big\rangle_{\rm SM} = 0,
\end{eqnarray}
fixing the value of the Fierz interference term $b_F = 0.1219(12)$ in
terms of the contribution of the ECVC effect~\cite{Ivanov2017a}.
Plugging Equation~(\ref{eq:16}) into Equation~(\ref{eq:14}), we obtain
\vspace{-6pt}
\begin{adjustwidth}{-\extralength}{0cm}
\begin{equation}
\begin{aligned}
\begin{array}{llll}\label{eq:18}
\hspace{-0.3in}&&a_{\rm eff}(E_e, \vec{k}_e\cdot \vec{k}_{\nu}) =\Big(a(E_e) - \frac{1}{1 +
  3\lambda^2}\,\frac{m^2_e \Delta^2}{q^4} + \frac{1}{1 +
  3\lambda^2}\frac{1}{2}\Big(|C_T|^2 + |\bar{C}_T|^2 -
    |C_S|^2 - |\bar{C}_S|^2\Big)\Big)/Z(E_e),\\
\hspace{-0.3in}&&A_{\rm eff}(E_e, \vec{k}_e\cdot \vec{k}_{\nu}) = \Big(A(E_e) - \frac{1}{1 +
      3\lambda^2}{\rm Re}\Big(2 C_T \bar{C}^*_T + (C_S
      \bar{C}^*_T + \bar{C}_S C^*_T)\Big)\Big)/Z(E_e),\\
\hspace{-0.3in}&&B_{\rm eff}(E_e, \vec{k}_e\cdot \vec{k}_{\nu}) =
\Big(B(E_e)- \frac{1}{2}\,(A_0 + B_0)\,\frac{m^2_e \Delta}{q^2 E_e} -
b_{ST}\,\frac{m_e}{E_e}- \frac{1}{1 + 3\lambda^2}{\rm Re}\Big(2 C_T
\bar{C}^*_T - (C_S \bar{C}^*_T + \bar{C}_S C^*_T)\Big)\Big)/Z(E_e),\\
\hspace{-0.3in}&&D_{\rm eff}(E_e, \vec{k}_e\cdot \vec{k}_{\nu}) =
\Big(D(E_e)- \frac{1}{2}\,(A_0 + B_0)\,\frac{ \alpha\, m^2_e
  \Delta}{q^2 k_e} + \frac{1}{1 + 3\lambda^2}\,\frac{1}{2}\,{\rm
  Im}\Big(C_S C^*_T + \bar{C}_S \bar{C}^*_T\Big)\Big)/Z(E_e),
\end{array}
\end{aligned}
\end{equation}
\end{adjustwidth}
with
 a common denominator
\begin{eqnarray}\label{eq:den18}
\hspace{-0.3in}&&Z(E_e)=1 - \frac{\Delta f_n}{f_n} - b_F\,\Big\langle\frac{m_e}{E_e}\Big\rangle_{\rm SM} + \Big[b_F -  \frac{1}{1 + 3\lambda^2}\,\Big(\frac{2 m_e\Delta }{q^2} - \frac{m_e    E_e\Delta^2}{q^4}\Big)\Big]\,\frac{m_e}{E_e}.
\end{eqnarray}

This is a complete set of correlation coefficients, which can be used
for a fit of the experimental data on the electron energy and angular
distributions of neutron $\beta^-$-decay with a polarized neutron
and unpolarized proton and electron.

Since our numerical analysis of the relative contributions of the ECVC
effect to the cross sections for the quasi-elastic electron
neutrino--neutron scattering and for the inverse $\beta$-decay,
carried out in Appendix~\ref{appendixa} (see Figure~\,\ref{fig:fig2}), shows that
these processes are not sensitive to the ECVC effect, the experimental
analysis of the correlation coefficients, given in Equation~(\ref{eq:18}),
is a matter of great importance for understanding whether the ECVC
hypothesis is correct in the sense that a matrix element of the
divergence of the charged weak hadronic vector current $\langle
h'|\partial^{\mu}V^{(+)}_{\mu}(0)|h\rangle$ vanishes, i.e., $\langle
h'|\partial^{\mu}V^{(+)}_{\mu}(0)|h\rangle = 0$, for the hadronic $h
\to h'$ transitions, even for different masses of incoming $h$ and
outgoing $h$ hadrons, or not. However, it is obvious that without
contributions of interactions beyond the SM, such a hypothesis cannot
be fulfilled.

If such an experimental analysis of the ECVC effect shows that  the correlation coefficients Equation~(\ref{eq:18}) are irrelevant, the contributions of the term $F_4(q^2) q_{\mu} \hat{q}/M^2$ to the neutron lifetime and correlation coefficients of neutron $\beta^-$-decay should be of order $O(m_e\Delta/M^2)$ or even smaller in comparison with the results obtained in~\cite{Ivanov2013,Ivanov2017b}.

In turn, in case of positive results of the experimental analysis of
the ECVC effect using the correlation coefficients Equation~(\ref{eq:18}), they should be taken
into account for the description of the quasi-elastic
neutrino--neutron scattering, the inverse $\beta$-decay, as well
as in the energy spectra of the neutrino production of nucleon
resonances. The values of the scalar and tensor phenomenological coupling constants
$C_S,\bar{C}_S, C_T$ and $\bar{C}_T$ together with the Fierz
interference term $b_F$ and the correlation coefficients $b_{ST}$,
which can be obtained by means of the fit of the experimental data on
different asymmetries of neutron $\beta^-$-decay.

\section{Conclusions}
We are examining the validity of the hypothesis regarding the precise conservation of the charged weak hadronic vector current in neutron beta-decay, utilizing polarized neutrons and unpolarized protons and electrons. Our analysis involves the computation of contributions from the phenomenological term, which is responsible for the Exact Conservation of the charged weak hadronic Vector Current (ECVC effect) in neutron beta-decay, even when considering varying masses of the neutron and proton.
We calculate correlation coefficients, incorporating the complete set of contributions from scalar and tensor interactions beyond the Standard Model (SM). We posit that if the cumulative effects of scalar and tensor interactions beyond the SM fail to reconcile with the experimental data on correlation coefficients alongside the contributions of the ECVC effect, it may be inferred that the conservation of the charged weak hadronic vector current is not maintained in the hadronic transitions of weak processes involving distinct masses of incoming and outgoing hadrons.

\vspace{6pt}

\authorcontributions{{D.A.: Conceptualization, Methodology, Data Curation, Formal analysis, Investigation, Writing—Original draft preparation. R.H.: Conceptualization, Formal analysis, Data Curation, Investigation, Methodology, Software, Validation, Writing—Original draft preparation. M.W.: Conceptualization, Formal analysis, Data Curation, Investigation, Methodology, Software, Validation, Writing—Original draft preparation. All authors contributed equally and agreed to the published version of the manuscript. The sole responsibility for the content of this publication lies with the authors.}} 

\funding{{The work of M. Wellenzohn was supported by MA 23 (p.n. 30-22).}} 

\dataavailability{{The data and illustrations presented in this study can be obtained directly from the equations. All data are available on request from the corresponding author.}} 

\acknowledgments{We want to thank our dear colleague  Andrey Nikolaevich Ivanov, who was the main investigator of this work until he sadly passed away on 18 December 2021. We see it as our professional and personal duty to honor his legacy by continuing to publish our collaborative work. Andrey was born on 3 June 1945 in what was then Leningrad. Since 1993 he was a university professor at the Faculty of Physics, named ``Peter The Great St. Petersburg Polytechnic University'' after Peter the Great. Since 1995 he has been a guest professor at the Institute for Nuclear Physics at the Vienna University of Technology for several years and has been closely associated with the institute ever since. This is also were we met Andrey and have been collaborating with him closely for more than 20 years, resulting in 40 scientific publications; see also~\cite{Ivanov:2013fca, Ivanov:2018qen, Ivanov:2018ngi, Ivanov:2018olo, Ivanov:2018yir, Ivanov:2019bqr, Ivanov:2020ybx, Ivanov:2021bae, Ivanov:2021lji, Ivanov:2021yhl, Ivanov:2021xkm}. We will miss Andrey as a personal friend and we will miss his immense wealth of ideas, scientific skills and his creativity. See also the official obituary (\url{https://www.tuwien.at/en/phy/ati/news/test}, accessed on 3 November 2024) for Andrey Nikolaevich Ivanov. The work of M. Wellenzohn was supported by MA 23 (p.n. 27-07 and p.n. 30-22). The sole responsibility for the content of this publication lies with the authors. }

\conflictsofinterest{{The authors declare no conflicts of interest.}} 

\appendixtitles{yes} 
\appendixstart
\appendix
\section[\appendixname~\thesection]{Cross Sections for the Quasi-Elastic Electron Neutrino--Neutron Scattering and the Inverse \boldmath{$\beta$}-Decay}
\renewcommand{\theequation}{A\arabic{equation}}
\setcounter{equation}{0}
\label{appendixa}

In this Appendix we calculate the cross sections for the
quasi-elastic scattering $\nu_e + n \to p + e^-$ and the inverse
$\beta$-decay $\bar{\nu}_e + p \to n + e^+$ by taking into account
the contributions of the ECVC effect. Neglecting the contributions of
the weak magnetism, recoil and radiative corrections and skipping
intermediate standard calculations, we obtain the following cross
sections for the quasi-elastic electron neutrino--neutron scattering
$\sigma(E_{\nu})$ and the inverse $\beta$-decay
$\sigma(E_{\bar{\nu}})$:
\begin{adjustwidth}{-\extralength}{0cm}
\begin{equation}
\begin{aligned}
\begin{array}{lll}\label{eq:A.1}
\sigma(E_{\nu}) &=& \sigma_0(E_{\nu}) +
\frac{G^2_F|V_{ud}|^2}{2\pi}\,k_- E_-\Big\{- \frac{m^2_e\Delta}{k_-
  E_- E_{\nu}}\Big[{\ell n}\Big(1 + \frac{2 k_-E_{\nu}\beta_- }{m^2_e
    - 2 E_- E_{\nu}}\Big) - {\ell n}\Big(1 - \frac{2 k_-E_{\nu}
  }{m^2_e - 2 E_- E_{\nu}}\Big)\Big]\\ &+&
\frac{2m^2_e\Delta^2}{(m^2_e - 2 E_- E_{\nu})^2 - 4 k^2_- E^2_{\nu}} -
\frac{m^2_e \Delta^2}{4 k_- E_- E^2_{\nu}}\Big[{\ell n}\Big(1 +
  \frac{2 k_-E_{\nu}\beta_- }{m^2_e - 2 E_- E_{\nu}}\Big) - {\ell
    n}\Big(1 - \frac{2 k_-E_{\nu} }{m^2_e - 2 E_-
    E_{\nu}}\Big)\\ &-&\frac{4 k_- E_{\nu}(m^2_e - 2 E_-
    E_{\nu}) }{(m^2_e - 2 E_- E_{\nu})^2 - 4 k^2_-
    E^2_{\nu}}\Big]\Big\},
\end{array}
\end{aligned}
\end{equation}
\end{adjustwidth}
where $E_- = E_{\nu} + \Delta$, $k_- = \sqrt{E^2_- - m^2_e}$ and
$\beta_- = k_-/E_-$ are the energy, momentum and velocity of the
electron, and
\begin{adjustwidth}{-\extralength}{0cm}
\begin{equation}
\begin{aligned}
\begin{array}{lll}\label{eq:A.2}
\sigma(E_{\bar{\nu}}) &=& \sigma_0(E_{\bar{\nu}}) +
\frac{G^2_F|V_{ud}|^2}{2\pi}\,k_+ E_+\Big\{\frac{m^2_e\Delta}{k_+ E_+
  E_{\bar{\nu}}}\Big[{\ell n}\Big(1 + \frac{2 k_+ E_{\bar{\nu}}
  }{m^2_e - 2 E_+ E_{\bar{\nu}}}\Big) - {\ell n}\Big(1 - \frac{2
    k_+ E_{\bar{\nu}} }{m^2_e - 2 E_+
    E_{\bar{\nu}}}\Big)\Big]\\ &+&
\frac{2m^2_e\Delta^2}{(m^2_e - 2 E_+ E_{\bar{\nu}})^2 - 4 k^2_+
  E^2_{\bar{\nu}}} - \frac{m^2_e \Delta^2}{4 k_+ E_+ E^2_{\bar{\nu}}}\Big[{\ell
    n}\Big(1 + \frac{2 k_+ E_{\bar{\nu}} }{m^2_e - 2 E_+ E_{\bar{\nu}}}\Big) -
  {\ell n}\Big(1 - \frac{2 k_+ E_{\bar{\nu}} }{m^2_e - 2 E_+
    E_{\bar{\nu}}}\Big)\\ &-&\frac{4 k_+ E_{\bar{\nu}}(m^2_e - 2 E_+
    E_{\bar{\nu}}) }{(m^2_e - 2 E_+ E_{\bar{\nu}})^2 - 4 k^2_+
    E^2_{\bar{\nu}}}\Big]\Big\},
\end{array}
\end{aligned}
\end{equation}
\end{adjustwidth}
where $E_+ = E_{\bar{\nu}} - \Delta$ and $k_+ = \sqrt{E^2_+ - m^2_e}$
are the energy and momentum of the positron. The cross sections $
\sigma_0(E_{\nu})$ and $\sigma_0(E_{\bar{\nu}})$ are given by
\cite{Ivanov2013a}
\begin{eqnarray}\label{eq:A.3}
\sigma_0(E_{\nu}) = (1 + 3 \lambda^2)\,
\frac{G^2_F|V_{ud}|^2}{\pi}\,k_- E_-\quad,\quad
\sigma_0(E_{\bar{\nu}}) = (1 + 3 \lambda^2)\,
\frac{G^2_F|V_{ud}|^2}{\pi}\,k_+ E_+.
\end{eqnarray}

\begin{figure}[H]

\begin{adjustwidth}{-\extralength}{0cm}
\centering 
\includegraphics[width=.65\textwidth]{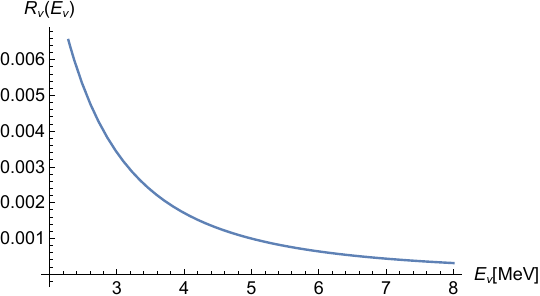}
\includegraphics[width=.65\textwidth]{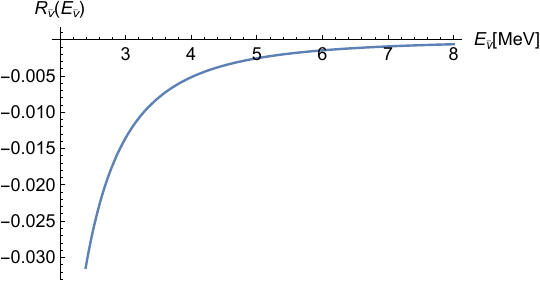}
\end{adjustwidth}
  \caption{The relative contributions $R_{\nu}(E_{\nu})$ (\textbf{left}) and
    $R_{\bar{\nu}}(E_{\bar{\nu}})$ (\textbf{right}) of the ECVC effect to the
    cross sections for the quasi-elastic electron neutrino--neutron
    and inverse $\beta$-decay in the neutrino and antineutrino energy
    regions $2\,{\rm MeV} \le E_{\nu} \le 8\,{\rm MeV}$ and $2\,{\rm
      MeV} \le E_{\bar{\nu}} \le 8\,{\rm MeV}$, calculated for
    $\lambda = - 1.2750$~\cite{Ivanov2013a}. }
\label{fig:fig2}
\end{figure}

In the quasi-elastic electron neutrino--neutron scattering and the
inverse $\beta$-decay the energies of neutrino and antineutrino vary
in the regions $E_{\nu} \ge 0$ and $E_{\bar{\nu}} \ge
(E_{\bar{\nu}})_{\rm thr} = ((m_n + m_e)^2 - m^2_p)/2m_p =
1.8061\,{\rm MeV}$~\cite{Ivanov2013a}.  The terms dependent on
$\Delta$ are caused by the ECVC effect. We define the relative contributions of
the ECVC effect to the cross sections under consideration as
follows
\vspace{-10pt}
\begin{adjustwidth}{-\extralength}{0cm}
\begin{eqnarray}\label{eq:A.4}
\hspace{-0.3in}&&R_{\nu}(E_{\nu}) = \frac{1}{2}\,\frac{1}{1 +
  3\lambda^2}\Big\{- \frac{m^2_e\Delta}{k_- E_- E_{\nu}}\Big[{\ell
    n}\Big(1 + \frac{2 k_-E_{\nu}\beta_- }{m^2_e - 2 E_- E_{\nu}}\Big)
  - {\ell n}\Big(1 - \frac{2 k_-E_{\nu} }{m^2_e - 2 E_-
    E_{\nu}}\Big)\Big]+ \frac{2m^2_e\Delta^2}{(m^2_e - 2 E_-
  E_{\nu})^2 - 4 k^2_- E^2_{\nu}}\nonumber\\
\hspace{-0.3in}&& - \frac{m^2_e
  \Delta^2}{4 k_- E_- E^2_{\nu}}\Big[{\ell n}\Big(1 + \frac{2
    k_-E_{\nu}\beta_- }{m^2_e - 2 E_- E_{\nu}}\Big) - {\ell n}\Big(1 -
  \frac{2 k_-E_{\nu} }{m^2_e - 2 E_-
    E_{\nu}}\Big) -\frac{4 k_- E_{\nu}(m^2_e - 2 E_-
    E_{\nu}) }{(m^2_e - 2 E_- E_{\nu})^2 - 4 k^2_-
    E^2_{\nu}}\Big]\Big\},
\end{eqnarray}
\end{adjustwidth}
and 
\vspace{-10pt}
\begin{adjustwidth}{-\extralength}{0cm}
\begin{eqnarray}\label{eq:A.5}
\hspace{-0.3in}&&R_{\bar{\nu}}(E_{\bar{\nu}}) =
\frac{1}{2}\,\frac{1}{1 + 3\lambda^2} \Big\{\frac{m^2_e\Delta}{k_+ E_+
  E_{\bar{\nu}}}\Big[{\ell n}\Big(1 + \frac{2 k_+ E_{\bar{\nu}}
  }{m^2_e - 2 E_+ E_{\bar{\nu}}}\Big) - {\ell n}\Big(1 - \frac{2 k_+
    E_{\bar{\nu}} }{m^2_e - 2 E_+ E_{\bar{\nu}}}\Big)\Big] +
\frac{2m^2_e\Delta^2}{(m^2_e - 2 E_+ E_{\bar{\nu}})^2 - 4 k^2_+
  E^2_{\bar{\nu}}}\nonumber\\
\hspace{-0.3in}&& - \frac{m^2_e \Delta^2}{4 k_+ E_+
  E^2_{\bar{\nu}}}\Big[{\ell n}\Big(1 + \frac{2 k_+ E_{\bar{\nu}}
  }{m^2_e - 2 E_+ E_{\bar{\nu}}}\Big) - {\ell n}\Big(1 - \frac{2 k_+
    E_{\bar{\nu}} }{m^2_e - 2 E_+ E_{\bar{\nu}}}\Big) - \frac{4 k_+
    E_{\bar{\nu}}(m^2_e - 2 E_+ E_{\bar{\nu}}) }{(m^2_e - 2 E_+
    E_{\bar{\nu}})^2 - 4 k^2_+ E^2_{\bar{\nu}}}\Big]\Big\},
\end{eqnarray}
\end{adjustwidth}
where $R_{\nu}(E_{\nu}) = \Delta \sigma(E_{\nu})/\sigma_0(E_{\nu})$,
$R_{\bar{\nu}}(E_{\bar{\nu}}) = \Delta
\sigma(E_{\bar{\nu}})/\sigma_0(E_{\bar{\nu}})$ with $\Delta
\sigma(E_{\nu}) = \sigma(E_{\nu}) - \sigma_0(E_{\nu})$ and $\Delta
\sigma(E_{\bar{\nu}}) = \sigma(E_{\bar{\nu}}) -
\sigma_0(E_{\bar{\nu}})$, respectively. The cross sections
Equations~(\ref{eq:A.1}) and~(\ref{eq:A.2}) are calculated in the
laboratory frame in the non-relativistic approximation for outgoing
hadrons. Since the most important region of the antineutrino energies
for the inverse $\beta$-decay is $2\,{\rm MeV} \le E_{\bar{\nu}} \le
8\,{\rm MeV}$~\cite{Ivanov2013a}, in Figure~\,\ref{fig:fig2}, we plot
$R_{\nu}(E_{\nu})$ and $R_{\bar{\nu}}(E_{\bar{\nu}})$ for $E_{\nu}$
and $E_{\bar{\nu}}$ varying over the regions $2\,{\rm MeV} \le E_{\nu}
\le 8\,{\rm MeV}$ and $2\,{\rm MeV} \le E_{\bar{\nu}} \le 8\,{\rm
  MeV}$,~respectively.

Our numerical examination of the relative impacts of the ECVC effect on the cross sections for both quasi-elastic electron neutrino–neutron scattering and inverse $\beta$-decay indicates that these processes exhibit low sensitivity to the ECVC effect. Specifically, the contribution of the ECVC effect to the cross section for quasi-elastic electron neutrino–neutron scattering is less than $0.7\,\%$ at $E_{\nu} \simeq 2\,{\rm MeV}$ and diminishes significantly by approximately two orders of magnitude at $E_{\nu} \simeq 8\,{\rm MeV}$.
The cross-section analysis for inverse $\beta$-decay, utilized in examining the deficit of positrons induced by reactor electron antineutrinos~\cite{Ivanov2013a,Mention2013}, should be averaged over the reactor electron antineutrino energy spectrum, which peaks at $E_{\bar{\nu}} \simeq 4\,{\rm MeV}$. According to Figure~\,\ref{fig:fig2}, the contribution of the ECVC effect is expected to decrease the yield of positrons $Y_{e^+}$ by about $0.5\,\%$. Given that this contribution is smaller than the experimental error bars $Y_{e^+} = 0.943(23)$~\cite{Mention2013}, one could argue that the inverse $\beta$-decay is insensitive to the influence of the ECVC effect.

\begin{adjustwidth}{-\extralength}{0cm}

\reftitle{References}

\PublishersNote{}
\end{adjustwidth}
\end{document}